\documentclass[final,5p,times,twocolumn]{elsarticle}

\usepackage{amssymb}
\usepackage{bm}% bold math
\usepackage[colorlinks=true,citecolor=magenta,urlcolor=cyan]{hyperref}
\usepackage{xspace}% bold math

\newcounter{bla}

\journal{Computer Physics Communications}
\def\ptv{\textbf{\textit{P}}_{\rm{T}}}

\def\pt{P_{\rm{T}}}

\def\STv{\textbf{\textit{S}}_{\rm{T}}}
\def\ST{\textbf{\textit{S}}_{\rm{T}}}

\def\Gevc{\rm{GeV}/\textit{c}}
\def\gevc2{\rm{GeV}/\textit{c}^2}
\def\Mhh{\textit{M}_{\rm{hh}}}

\def\Acoll{A_{\rm{UT}}^{\sin(\phih+\phi_{\rm S}-\pi)}}

\def\phih{\phi_{\rm{h}}}
\def\phiS{\phi_{\rm{S}}}
\def\phiC{\phi_{\rm{C}}}
\def\phiRS{\phi_{\rm{RS}}}
\def\phiR{\phi_{\rm R}}

\def\Dnn{D_{\rm{NN}}}
\def\Ahh{A_{\rm{UT}}^{\sin\phiRS}}
\def\RT{\textbf{R}_{\rm T}}
\def\ptOne{\textbf{P}_{\rm{1T}}}
\def\ptTwo{\textbf{P}_{\rm{2T}}}

\def\xu{\hat{\textbf{x}}}

\def\zu{\hat{\textbf{z}}}

\def\pythia{\textsc{Pythia}\xspace}
\def\3p0{string+${}^3P_0$}
\def\StringSpinner{\texttt{StringSpinner}\xspace}
\def\kpt{\textbf{k}'_{\rm{T}}}
\def\kptkpt{\textbf{k}'^2_{\rm{T}}}
\def\kptabs{\rm{k}'_{\rm{T}}}
\def\kpthat{\hat{\textbf{k}}'_{\rm T}}
\def\SqT{\textbf{S}_{q\rm{T}}}
\def\SqL{S_{q\rm{L}}}
\def\Sq{\textbf{S}_{q}}
\def\Im{\rm{Im}}
\def\fL{f_{\rm{L}}}
\def\GL{G_{\rm{L}}}
\def\GT{G_{\rm{T}}}
\def\thetaLT{\theta_{\rm{LT}}}
\def\D{D}
\def\vecsigma{\boldsymbol{\sigma}}

\def\eg{\textit{e.g.}}
\def\ie{\textit{i.e.}}

\def\setting#1{{\small\texttt{ #1}}}
\def\settingval#1#2{{\small\texttt{ #1 = } #2}}

\begin{document}

\begin{frontmatter}



%\title{Inclusion of quark-spin effects for vector meson production in \texttt{PYTHIA} hadronization}
%\title{Implementation of quark spin effects for vector meson production in deep-inelastic scattering in PYTHIA with StringSpinner}
% \title{Extension of StringSpinner to the inclusion of vector mesons and their decays \\ in string fragmentation \\
% \red{(not that simple and nice title)}}
%\title{Simulation of spin effects in semi-inclusive deep-inelastic scattering with PYTHIA and description of the transverse-spin asymmetry data}
\title{Extending StringSpinner to handle vector-meson spin}

%A \author[a]{A. Kerbizi\corref{author}}
\author[a]{Albi Kerbizi\corref{author}}
\author[b]{Leif L\"onnblad}
%\author[b]{Third Author}

\cortext[author] {Corresponding author.\\\textit{E-mail address:} albi.kerbizi@ts.infn.it}
\address[a]{Dipartimento di Fisica, Universit\'a degli Studi di Trieste and INFN Sezione di Trieste\\
  Via Valerio 2, 34127 Trieste, Italy}
\address[b]{Department of Physics, Box 118, 221 00 LUND, Sweden}

%\cortext[author] {Corresponding author.\\\textit{E-mail address:} }

\begin{abstract}
  Quark spin effects in hadronization were recently included in the
  \pythia~8 Monte Carlo event generator for the simulation of the deep
  inelastic scattering (DIS) process off a polarized proton or neutron
  target. The spin effects were activated via the external
  \StringSpinner package, which is based on the string+${}^3P_0$ model
  of polarized hadronization, and were restricted to the production of
  pseudoscalar mesons in string fragmentation. The struck quark
  polarization could either be calculated using parametrizations of the
  transversity PDFs or be freely specified.
  In this article we present a major extension of \StringSpinner that
  includes vector meson production in the \pythia~8 string
  fragmentation. The spin-dependent emission and decay processes of
  the vector mesons are described using the string+${}^3P_0$ model,
  which systematically accounts for the quark-spin degree of freedom
  in the fragmentation chain. The new package works with the version
  8.3 of the \pythia event generator and it allows for a more complete
  and quantitative simulation of the final states in polarized
  semi-inclusive DIS events. Also, it can be used to study the
  transverse-spin effects for the production of vector mesons. After
  describing the new version of \StringSpinner, examples of its usage
  in semi-inclusive DIS events are presented.

%A submitted program is expected to satisfy the following criteria: it must be of benefit to other physicists, or be an exemplar of good programming practice, or illustrate new or novel programming techniques which are of importance to computational physics community; it should be implemented in a language and executable on hardware that is widely available and well documented; it should meet accepted standards for scientific programming; it should be adequately documented and, where appropriate, supplied with a separate User Manual, which together with the manuscript should make clear the structure, functionality, installation, and operation of the program.

%Your manuscript and figure sources should be submitted through Editorial Manager (EM) by using the online submission tool at \\
%https://www.editorialmanager.com/comphy/.

%In addition to the manuscript you must supply: the program source code; a README file giving the names and a brief description of the files/directory structure that make up the package and clear instructions on the installation and execution of the program; sample input and output data for at least one comprehensive test run; and, where appropriate, a user manual.

%A compressed archive program file or files, containing these items, should be uploaded at the "Attach Files" stage of the EM submission.

%For files larger than 1Gb, if difficulties are encountered during upload the author should contact the Technical Editor at cpc.mendeley@gmail.com.

\end{abstract}


\end{frontmatter}

% \linenumbers

% All CPiP articles must contain the following
% PROGRAM SUMMARY.

{\bf NEW VERSION PROGRAM SUMMARY}
  %Delete as appropriate.

\begin{small}
\noindent
{\em Program Title: StringSpinner}                                          \\
{\em CPC Library link to program files: } (to be added by Technical Editor) \\
{\em Developer's repository link:\\
   \href{https://gitlab.com/albikerbizi/stringspinner.git}{~~~~https://gitlab.com/albikerbizi/stringspinner.git}} \\
{\em Code Ocean capsule:} %(to be added by Technical Editor)
\\
{\em Licensing provisions: GNU GPL v2 or later}  \\
{\em Programming language: \texttt{C++}, \texttt{Fortran}}    \\
%{\em Supplementary material:}                                 \\
  % Fill in if necessary, otherwise leave out.
{\em Journal reference of previous version: }*                  \\
  %Only required for a New Version summary, otherwise leave out.
{\em Does the new version supersede the previous version?: Yes}   \\
  %Only required for a New Version summary, otherwise leave out.
{\em Reasons for the new version: Vector meson emission and decay is required for a more complete simulation of the DIS final states.}\\
  %Only required for a New Version summary, otherwise leave out.
{\em Summary of revisions: The spin-dependence of the string fragmentation is extended to include the production and the decay of polarized vector mesons.}\\
  %Only required for a New Version summary, otherwise leave out.
{\em Nature of problem: Vector meson contribute strongly to the final hadrons produced in DIS, and they can produce non-trivial effects on the spin-dependent observables.}\\
  %Describe the nature of the problem here. \\
{\em Solution method: Extend the previous version of \setting{StringSpinner} to include polarized vector meson production and decay using a more complete model of polarized hadronization.}\\
  %Describe the method solution here.
%{\em Additional comments including restrictions and unusual features (approx. 50-250 words):}\\
  %Provide any additional comments here.
   \\

\end{small}

\section{Introduction} \label{sec:Introduction} The quark spin effects
were recently introduced in the hadronization part of the \pythia~8.2
Monte Carlo event generator (MCEG) \cite{Sjostrand:2014zea} for the
simulation of the deep inelastic scattering (DIS) process via the
\StringSpinner package \cite{Kerbizi:2021StringSpinner}. The main
purpose was the simulation of the semi-inclusive DIS (SIDIS) process
off a transversely polarized proton or neutron, and in
particular the study of the transverse-spin effects induced by the
transversity parton distribution function (PDF) $h_1^q$. The
transversity PDF gives the transverse polarization of a quark $q$ in a
transversely polarized nucleon and together with the spin-averaged PDF
$f_1^q$ and helicity PDF $g_1^q$ completes the collinear partonic
structure of the nucleons at leading order.

Among the observable effects originated by $h_1^q$ in SIDIS there are the Collins and dihadron transverse-spin asymmetries (TSAs). In a factorized approach of SIDIS, such asymmetries arise from the coupling between $h_1^q$ and a spin-dependent fragmentation function (FF), and show up as the amplitudes of specific modulations in the azimuthal distributions of the final state hadrons. If single final state hadrons are considered the related FF is the Collins function $H_{1q}^{\perp h}$ \cite{Collins:1992kk} that describes the fragmentation of a transversely polarized quark in unpolarized hadrons, and the corresponding TSA is the Collins asymmetry \cite{Collins:1992kk}. If pairs of final state hadrons are considered the corresponding FF is the interference fragmentation function $H_{1q}^{\sphericalangle\,hh}$ that describes the fragmentation of a transversely polarized quark in a pair of unpolarized hadrons, and the related TSA is the dihadron asymmetry \cite{Bianconi:1999cd}. The Collins and the dihadron asymmetries have been measured by different experiments \cite{HERMES:2010mmo,COMPASS:2014bze,COMPASS:2014ysd}, and have been used in combination with the corresponding $e^+e^-$ data for the extraction of $h_1^q$ and the spin-dependent FFs $H_{1q}^{\perp\, h}$ and $H_{1q}^{\sphericalangle\, hh}$. In another observable, $h_1^q$ is coupled to a jet-function which describes the conversion of a transversely polarized quark in a jet of hadrons \cite{Lai:2022aly}. Unlike the Collins and dihadron asymmetries, this observable presently has not been measured. For a review on transverse-spin effects in hard semi-inclusive collisions see, \eg, Ref. \cite{Anselmino:2020vlp}.

The main ingredients introduced in \StringSpinner to simulate the polarized SIDIS process are the parametrizations of $h_1^q$ \cite{Martin:2014wua} and the string+${}^3P_0$ model of polarized hadronization \cite{Kerbizi:2019ubp}. The parametrizations of $h_1^q$ are implemented for valence quarks and are used to calculate the transverse polarization of the quarks entering the hard scattering. The string+${}^3P_0$ model is used for the simulation of the string fragmentation process of the polarized quarks \cite{Kerbizi:2018qpp}. Since the quark spin degree of freedom is propagated along the full fragmentation chain in a systematic way, the string+${}^3P_0$ model is expected to reproduce not only the Collins and dihadron asymmetries but also possible spin effects in more exclusive hadronic final states, \eg, as those described by the jet functions.

In the first step toward the end goal of a systematic introduction of the quark-spin effects in the \pythia event generator, some restrictions were applied in the \StringSpinner version of Ref. \cite{Kerbizi:2021StringSpinner}. They were essentially due to the current development of the string+${}^3P_0$ model.
The string fragmentation process was in fact restricted to the production of pseudoscalar (PS) mesons, while other hadronic states such as the vector mesons (VMs) were ignored\footnote{We also recall that in \StringSpinner the parton showers are switched off because presently the string+${}^3P_0$ model does not handle the more general string configurations involving multiple partons that would be produced in the showering process.}. It is however known that a relevant fraction of the observed final state hadrons consists in the decay products of VMs. These secondary mesons contribute, together with the hadrons directly produced in the string fragmentation, in building the spin-dependent observables in SIDIS. Understanding the contribution of VMs is thus important for the interpretation of the data. Moreover, transverse-spin effects in the production of VMs are interesting phenomena to explore as they provide a deep insight in the spin-dependence of the hadronization process \cite{Bacchetta:2000jk}. From the experimental point of view, data on transverse-spin effects for VMs is rather poor. The first measurement of the Collins asymmetries for $\rho^0$ mesons was recently performed by the COMPASS experiment \cite{COMPASS:2022jth}.

In this article we present a major extension of \StringSpinner that includes the production and the decays of VMs in the polarized string fragmentation process of the \pythia event generator. For this purpose, we use the recently developed string+${}^3P_0$ model with PS meson and VM production \cite{Kerbizi:2021M20}. The model introduces two new free parameters to describe the relevant polarization states of the VMs. The decays of the VMs are simulated taking into account their polarization states. The spin information is propagated along the fragmentation chain using the Collins-Knowles (CK) recipe \cite{Collins:1987cp,Knowles:1988vs} to preserve the quantum mechanical correlations between the angular orientations of the decay hadrons and the polarization of the fragmenting quarks in the recursive fragmentation chain. Concerning the actual implementation, the \setting{UserHooks} class of \StringSpinner that enables external intervention in the \pythia string fragmentation process is extended to include the production of light and heavy VMs. The \setting{UserHooks} class is also updated to work with the version 8.3 of the \pythia generator \cite{Bierlich:2022pfr}. A new implementation of the \pythia class for external intervention in decay processes is developed in order handle the polarized hadronic decays of the VMs. This new development %allows for a more complete simulation of the final states and transverse spin effects in polarized DIS events, particularly those related to the inclusive vector meson production.
also provides the framework for possible future extensions of the \StringSpinner package aimed at introducing other species of hadrons and their decays.

The article is organized as follows. The implementation of the new quark spin effects in the \pythia string fragmentation and the handling of the external decays of VMs are presented in Sec. \ref{sec:extension}. The description of the program files and the instructions to use the new version of the \StringSpinner package are given in Sec. \ref{sec:program files}. The main simulation results of TSAs obtained with the new version of the package and the comparison with data are shown in Sec. \ref{sec:results}. The conclusions and the future prospects are discussed in Sec. \ref{sec:conclusions}.

\section{Extension of the polarized string fragmentation of \pythia}\label{sec:extension}
%The introduction of quark spin effects in the string fragmentation process of \pythia{} is achieved using the \texttt{UserHooks} class of the generator. This class allows an external user to step in during the execution of the normal string fragmentation process and to veto each produced hadron according to some externally imposed logic. The logic implemented in \StringSpinner is inspired to the \3p0{} model presented in Ref. \cite{Kerbizi:2021} and it is the following.

%The introduction of quark spin effects in the string fragmentation process of \pythia{} is achieved using the most recent string+${}^3P_0$ model presented in Ref. \cite{Kerbizi:2021gos}. The model includes

When a DIS event is simulated, \pythia generates the values of the Bjorken variable $x$ and of the virtuality $Q^2$ of the exchanged boson, which here is considered to be a virtual photon. Using the PDFs, this allows to select the flavor of the quark which undergoes the hard scattering with the beam lepton. The hard scattering is simulated and the four-momenta before and after the scattering are set up. As a next step, \pythia constructs the remnants of the target nucleon and ties a string between the struck quark $q$ and a remnant $\bar{q}_R$. The remnant $\bar{q}_R$ can be either a diquark or an antiquark depending on whether the struck quark comes from the valence or sea regions.

Before the fragmentation of the $q\bar{q}_R$ string is started, \StringSpinner calculates the density matrix $\rho(q)=(\textbf{1}+\boldsymbol{\sigma}\cdot\Sq)/2$ of the scattered quark either by using the parametrizations of the transversity PDFs or a value of the quark polarization chosen by the user, as described in Ref. \cite{Kerbizi:2021StringSpinner}. The vector $\Sq=(\SqT,\SqL)$ indicates the polarization vector of the scattered quark with transverse component $\SqT$ and longitudinal component $\SqL$ with respect to the string axis, which defines the longitudinal direction or $\zu$ axis. If the primordial transverse momentum of quarks is neglected, the string axis coincides with the momentum of the exchanged boson. The quantity $\boldsymbol{\sigma}=(\sigma_x,\sigma_y,\sigma_z)$ is a vector of Pauli matrices. The density matrix of the scattered quark is the initial condition for the simulation of the polarized string fragmentation process.

In the following subsections we describe only the modifications to \StringSpinner regarding the simulation of the spin effects in hadronization for the emission of PS mesons and VMs, and the simulation of the decays of VMs. The other features of the package are unchanged and the complete description of them is given in Ref. \cite{Kerbizi:2021StringSpinner}.

\subsection{Reweighting of hadrons in string fragmentation}\label{sec:reweight}
\begin{figure}[tbh]
\centering
\begin{minipage}[b]{0.5\textwidth}
\hspace{-0.8em}
\includegraphics[width=1.0\textwidth]{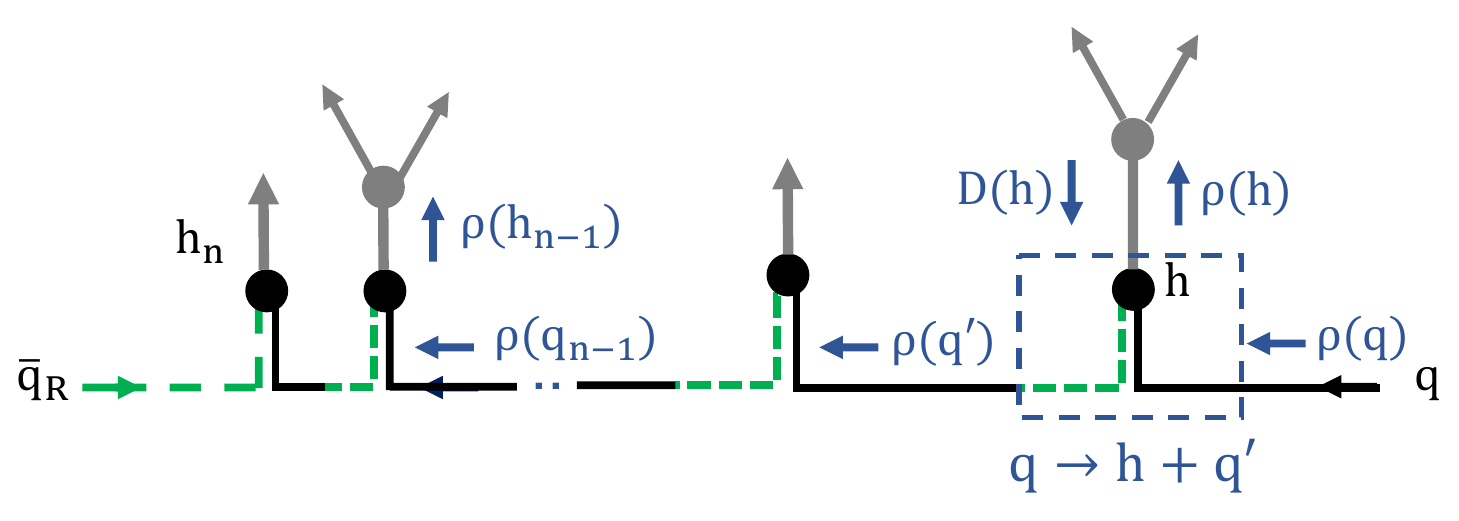}
\end{minipage}
\caption{Propagation of the spin information along the string fragmentation chain according to the recipe in Refs. \cite{Collins:1987cp,Knowles:1988vs}.}
\label{fig:spin propagation}
\end{figure}

 The string fragmentation process $q\,\bar{q}_R\rightarrow h_1,h_2,\dots$, where $h_1,h_2,\dots$ are the emitted hadrons, is simulated by \pythia as a recursive process of elementary quark splittings $q\rightarrow h + q'$ as shown in Fig. \ref{fig:spin propagation}. In the splitting the fragmenting quark $q$ emits the hadron $h$ and leaves the recurring quark $q'$ to be fragmented in the next splitting. The four momentum is conserved, \ie, it is $k = p + k'$, where $k$, $p$ and $k'$ indicate the four-momenta of $q$, $h$ and $q'$, respectively.

Each emitted hadron in string fragmentation is inspected by \StringSpinner, which applies a veto procedure to emulate the rules of the string+${}^3P_0$ model inspired by the standalone implementation of the model in Ref. \cite{Kerbizi:2021M20}. The hadron $h$ is rejected if it does not belong to the mupltiplets of (light or heavy) PS mesons or VMs, or if it is not produced from the fragmenting quark side. The latter rejection is such that the string fragmentation evolves from the fragmenting quark side towards the remnant side. If these selections are passed, the hadron is rejected with a probability $1/2$ in order not to change the composition of the produced hadrons but only their transverse momenta.

If the hadron passes these preliminary selections, it is accepted with the probability
\begin{equation}\label{eq:probability}
    p(\kpt,\SqT) = \frac{1}{2} \times \left[ 1 + a\, \frac{2\Im(\mu)\,k'_{\rm T}}{|\mu|^2+\kptkpt} \, \SqT\cdot \left(\zu\times \kpthat\right) \right],
\end{equation}
which is inspired by the polarized splitting probability of the string+${}^3P_0$ model in Ref. \cite{Kerbizi:2021M20}.
The vector $\kpt$ is the transverse momentum of $q'$ with respect to the string axis and $\kpthat=\kpt/\kptabs$. The mixed product $\SqT\cdot \left(\zu\times \kpthat\right)$ which appears in Eq. (\ref{eq:probability}) is responsible for the Collins effect in the fragmentation of transversely polarized quarks. The quantity $\mu$ is a complex mass parameter introduced in the \3p0{} model to parametrize the relative ${}^3P_0$ wave function of the $q'\bar{q}'$ pair produced at the string breaking. It is taken to be the same for all quark flavors. The factor $a$ is new with respect to Ref. \cite{Kerbizi:2021StringSpinner}. It is a constant given by $a=-1$ if $h$ is a PS meson and by $a=\fL$ if $h$ is a VM, where $\fL$ is a free parameter of the \3p0{} model governing the fraction of longitudinally polarized VMs. It is defined as $\fL=|\GL|^2/(2|\GT|^2+|\GL|^2)$, where $\GL$ and $\GT$ are the complex constants describing the coupling of quarks to VMs with longitudinal and transverse polarization with respect to the string axis, respectively.

As can be seen in Eq. (\ref{eq:probability}), the imaginary part of $\mu$ is responsible for the transverse spin effects in the fragmentation process, which vanish for $\rm{Im}(\mu)=0$. The factor $2\rm{Im}(\mu)k'_{\rm T}/(|\mu|^2+\kptkpt)$ regulates the magnitude of these effects and it is common to the multiplets of hadrons considered. The parameter $\fL$ governs mainly the spin effects for the production of VMs. Note also that it is $a<0$ for PS mesons and $a\geq 0$ for VMs. Hence the spin effects for PS mesons and VMs are opposite.

The real part of $\mu$ is instead responsible for the longitudinal spin effects as discussed in Ref. \cite{Kerbizi:2018qpp}.

\subsection{Decays of polarized vector mesons}\label{sec:decays}
In the standard procedure of \pythia, the decays of resonances produced during string fragmentation are performed after the termination of the fragmentation chain. In order to implement the CK recipe, however, it is necessary to perform the decays of the resonances right after the resonances are produced, as shown in Fig. \ref{fig:spin propagation}. In the figure, the short arrows indicate the propagation of the spin information along the chain. %This allows to propagate the quantum mechanical correlations between the decay products and the spin of the quark $q'$.

To implement this recipe, each accepted hadron $h$ is further inspected by \StringSpinner. If $h$ is a PS meson, its possible decay process is handled by \pythia as the meson does not carry any spin information. If $h$ is a VM it is decayed externally. The external decays are performed by the class named \setting{PolarizedDecayHandler}, which is new with respect to the previous version of \StringSpinner. It is a derived class of the \setting{DecayHandler} class offered by \pythia as a base class for the external handling of decays. A \setting{PolarizedDecayHandler} object is created inside the main \setting{UserHooks} class of \StringSpinner, and a pointer to this object is given to the main \pythia class together with the list of the hadrons to be decayed externally. This activates the external decays.

The actual decay process is performed within the \setting{PolarizedDecayHandler} class using the routines developed for the standalone implementation of the string+${}^3P_0$ model \cite{Kerbizi:2021M20}. Presently the VMs admitted for the external decays are the light mesons $\rho$ and $\omega$, and the strange mesons $\phi$ and $K^{*}$. The decay channels which can be treated externally are the hadronic two-body decays of the type VM$\rightarrow$ PS+PS and VM $\rightarrow$ PS+$\gamma$, and the three-body decays of the type VM$\rightarrow$ PS+PS+PS of the $\omega$ and $\phi$ mesons. Other decays are either not relevant for the goal of this work or have small branching ratios, and they are handled by the standard routines of \pythia.

To decay the hadron $h$ the required ingredients are the identity, the four-momentum in the string rest frame, and the polarization vector $\Sq$ of the fragmenting quark $q$. Knowing the identity of the VM, the decay channel is selected by the same functions as those used by \pythia in order to limit the introduction of external information, \eg, the mass, width and branching ratios. The momenta of the decay hadrons are generated using the spin density matrix $\rho(h)$ of $h$ and the same routines of the standalone implementation of the string+${}^3P_0$ model \cite{Kerbizi:2021M20}.

The density matrix of the VM is calculated as
\begin{eqnarray}\label{eq:rho(h)}
\hspace{-1.5em}\nonumber \rho_{\alpha\alpha'}(h) = \frac{ \rm{Tr} \left[(\mu + \sigma_z\, \vecsigma\!\cdot\!\kpt) \,\Gamma_{h,\alpha} \, \rho(q) \,
\Gamma^\dag_{h,\alpha'} \, (\mu^* +  \vecsigma\!\cdot\!\kpt \, \sigma_z) \right] } { \rm{Tr} \left[(\mu + \sigma_z\, \vecsigma\!\cdot\!\kpt) \,\Gamma_{h,\beta} \, \rho(q) \,
\Gamma^\dag_{h,\beta} \, (\mu^* +  \vecsigma\,\cdot\,\kpt \, \sigma_z) \right] },\\
\end{eqnarray}
where the trace is taken over the quark-spin indices, and a summation over the repeated indices is understood. The matrix $\Gamma_{h,\alpha} = \left(\GT\sigma_x\sigma_z, \GT\sigma_y\sigma_z,\GL\,1_{2\times 2}\right)$ is the coupling of the quarks $q$ and $q'$ to the VM with linear polarization along
$\alpha=x,y,z$. The density matrix depends on $\mu$, $\kpt$, $\fL$
and on the new parameter $\thetaLT$. The latter is defined as $\thetaLT=\arg(\GL/\GT)$ and it is the second free parameter of the \3p0{} model needed to describe the spin effects for VM production. A non-zero $\thetaLT$ is responsible for the oblique polarization of VMs, which is a new source of Collins effect as described in detail in Ref. \cite{Kerbizi:2021M20}.

The calculated density matrix $\rho(h)$ is used to generate the non-isotropic angular distribution of the decay products in the rest frame of the VM. They are then are boosted to the string rest frame, stored externally and provided to \pythia at a later stage as described below. The detailed explanation of the generation of the angular distribution of the decay products is given in Ref. \cite{Kerbizi:2021M20}.

%book-kept. For a more detailed explanation of the reached from the string rest frame by a sequence of a longitudinal and a transverse boost. polarized hadronic decays of the VM as explained in Ref. \cite{Kerbizi:2021} using as external routines those developed in the stand alone implementation of the \3p0{} model.

%The decay products of the VM are book-kept and provided to \pythia at at later stage (see below).

\subsection{Propagation of the spin information}\label{sec:spin propagation}
The external decays provide as a result the decay matrix $\D$ that is used to propagate the spin information along the string fragmentation chain, as shown in Fig. \ref{fig:spin propagation}. This matrix encodes the correlations between the polarization of the fragmenting quark $q$ and the decay products of the VM, as required by the CK recipe. It depends on the decay amplitude that describes the decay process. The corresponding expressions for the decays considered in this work can be found in Ref. \cite{Kerbizi:2021M20}.

The decay matrix is also used to calculate the density matrix of the leftover quark $q'$ \cite{Kerbizi:2021M20}
\begin{eqnarray}\label{eq:rho(q')}
   \nonumber
  \hspace{-4.5mm}\rho(q')\hspace{-2mm} &=& \hspace{-2mm}\frac{\left[
               \D_{\alpha'\alpha}\,(\mu + \sigma_z\, \vecsigma\cdot\kpt) \, \Gamma_{h,\alpha} \,
               \rho(q) \,  \Gamma^\dag_{h,\alpha'} \,
               (\mu^* +  \vecsigma\cdot\kpt \, \sigma_z)\right]}
               {\rm{Tr}\left[ Numerator \right]}. \\
\hspace{-2cm}
\end{eqnarray}
This expression ensures that the information about the orientation of the decay products is transferred to $q'$. In this way the spin information is propagated along the fragmentation chain accounting for quantum mechanical correlations.

\subsection{Closure of string fragmentation and storage of decay hadrons}
The steps described in the subsections \ref{sec:reweight}-\ref{sec:spin propagation} are repeated at each splitting generated during the string fragmentation process until the fragmentation chain is closed by the standard \pythia procedure (see Fig. \ref{fig:spin propagation}). To close the chain, the remaining string piece $q_{n-1}\bar{q}_R$, where $q_{n-1}$ is the leftover quark after $n-1$ hadrons have been emitted, is fragmented by \pythia generating a further string breaking where the $q_n\bar{q}_n$ pair is produced. No external veto procedure is performed on the construction of the final two hadrons, $h_{n-1}=(q_{n-1}\bar{q}_n)$ and $h_n=(q_{n}\bar{q}_R)$. However, if the hadron $h_{n-1}$ is a VM it is decayed externally as described in Sec. \ref{sec:decays}.

After the string fragmentation is terminated, the decays of the produced hadrons are simulated by \pythia except those which are deemed to be handled externally. The relevant information on the decays simulated externally is in fact transferred by the \setting{PolarizedDecayHandler} class to \pythia. The decay products of the externally decayed hadrons are thus stored in the event record and can be accessed by the standard \pythia methods. Their momenta are expressed in the same reference system as the one chosen by the user for the simulation of the DIS process.

\section{Program files and instructions}\label{sec:program files}
The new \StringSpinner package is composed of the old \texttt{C++}
files \texttt{StringSpinner.h}, \texttt{Transversity.h}, and
\texttt{dis.cc}; the old \texttt{Fortran} file \texttt{mc3P0.f90}; and
of the new \texttt{C++} files \texttt{VectorMesonDecays.h} and \texttt{PrimordialKT.h}, and
\texttt{Fortran} file \texttt{definitions.f90}. Among the old files,
\texttt{Transversity.h}, which implements the definitions of the
transversity PDFs, has not been modified. The description of the
modifications of the old files as well as the description of the new
ones is the following.

\texttt{StringSpinner.h} is the main file of the package containing the implementation of the \setting{UserHooks} class for the introduction of the spin effects in \pythia. It is updated
as compared to Ref. \cite{Kerbizi:2021StringSpinner} with modifications that allow to extend the list of hadrons that can be handled in string fragmentation according to the string+${}^3P_0$ model, and to decide which hadron is to be decayed externally. Also, the interface with \pythia is updated to work with the version 8.3 of the generator.

\setting{VectorMesonDecays.h} (new compared to Ref. \cite{Kerbizi:2021StringSpinner}) is a header file containing the \setting{PolarizedDecayHandler} class to manage the simulation of the polarized decay processes. It is an implementation of the \setting{ExternalDecays} class, offered by \pythia to allow the intervention of the external user during the decays. \setting{PolarizedDecayHandler} decides which decays can be performed externally, performs the decays by using the \texttt{Fortran} routines of \texttt{mc3P0.f90} (see below) and communicates with \pythia for the storage of the decay hadrons in the event record. This last step is performed by the other two classes named \setting{Decayer} and \setting{SavedHadrons}. These are new classes implemented to trace where the decay was performed during string fragmentation and to correctly save the decay products in the event record.

\setting{PrimordialKT.h} (new compared to Ref. \cite{Kerbizi:2021StringSpinner}) is a header file implementing an auxiliary class that allows to obtain the string axis in an event where the primordial transverse momentum of quarks is switched on.

\texttt{mc3P0.f90} implements the \texttt{Fortran} module that contains all the routines needed for the calculation of the spin-dependent quantities, \eg, the probability in Eq. (\ref{eq:probability}), the density matrices in Eq. (\ref{eq:rho(h)}) and Eq. (\ref{eq:rho(q')}), and for the generation of the angular distributions of the hadrons produced in the decay processes of the polarized VMs. This file is extended compared to Ref. \cite{Kerbizi:2021StringSpinner} and gathers the routines used in the standalone implementation of the string+${}^3P_0$ model in Ref. \cite{Kerbizi:2021M20}.

\texttt{definitions.f90} (new compared to Ref. \cite{Kerbizi:2021StringSpinner}) implements a \texttt{Fortran} module that contains the definitions of new types of variables. In particular it includes the definitions of four-vectors and spin density matrices as well as the definitions of the possible operations on these variables. The module is used only by \texttt{mc3P0.f90}.

The complete \StringSpinner package can be downloaded from
\texttt{gitlab}\footnote{\href{https://gitlab.com/albikerbizi/stringspinner.git}{https://gitlab.com/albikerbizi/stringspinner.git}.}. It
includes also a configure script and a \texttt{Makefile} for the
compilation of the main program, as will be explained in more detail
in Sec. \ref{sec:execution}.

\subsection{The main program}
As in Ref. \cite{Kerbizi:2021StringSpinner}, a sample program for the simulation of DIS events is provided by the \texttt{dis.cc} file. The structure of the main program is the one of standard \pythia with few additions, which are summarized as follows.

The schematic structure of the main program is
\small
\begin{verbatim}
#include "Pythia8/Pythia.h"
#include "StringSpinner.h"
using namespace Pythia8;
int main() {
  Pythia pythia;
  Event& event = pythia.event;
  auto fhooks =
      std::make_shared<SimpleStringSpinner>();
  fhooks->plugInto(pythia);

  // Standard Pythia settings.
  ....
  // StringSpinner settings.
  ....
  pythia.init(),
  for(int iEvent=0;iEvent<nEvents;iEvent++){
    if(!pythia.next()) continue;
    // Analysis of the Pythia event record using
    // the standard tools.
    ....
  }
  ....
  return 0;
}
\end{verbatim}
\normalsize The \texttt{StringSpinner.h} file must be included in the
main program. To activate the spin effects, the creation of the
\setting{Pythia} object by \setting{Pythia pythia}, must be followed by
the creation of a shared pointer to a \setting{SimpleStringSpinner}
object. As described in Ref. \cite{Kerbizi:2021StringSpinner},
\setting{SimpleStringSpinner} is the ad hoc implementation of the
\setting{UserHooks} class for the insertion of spin effects in
\pythia. The pointer is named here \setting{fhooks} and it can be
created by adding the line \setting{auto fhooks =
  std::make\_shared<SimpleStringSpinner>()}. The following command
\setting{fhooks->plugInto(pythia)} is needed to pass the pointer to the
created \setting{UsersHooks} class to the \setting{Pythia} class. This
allows \pythia to simulate the spin effects for the processes that can
be handled. It will also allow \StringSpinner to communicate with the
user using the \pythia settings system.

To change the settings of the generator, the standard \pythia
command \setting{pythia.readString()} can be used before the
\setting{pythia.init()} command. The same command can be used also for
the settings of \StringSpinner, at variance with the old version
where it was necessary to use ad hoc methods implemented in the
\setting{SimpleStringSpinner} class
\cite{Kerbizi:2021StringSpinner}. The \setting{pythia.readString()}
command takes a character string of the form \setting{"Parameter =
  value"} as argument, and in the following we will describe the
parameters available in \StringSpinner.

The target polarisation can be specified by the setting
\begin{itemize}
\item \settingval{StringSpinner:targetPolarisation}{\textit{Sx, Sy, Sz}}
\end{itemize}
where \textit{Sx}, \textit{Sy} and \textit{Sz} are the components of the polarisation vector in the chosen reference system. To chose freely the polarisation vector for a given quark flavor the analogue setting
\begin{itemize}
\item \settingval{StringSpinner:qPolarisation}{\textit{Sx, Sy, Sz}}
\end{itemize}
can be used, where \texttt{q} indicates the flavor of the quark and it can be \texttt{u}, \texttt{d}, \texttt{s}, \texttt{c}, \texttt{b}, \texttt{ubar}, \texttt{dbar}, \texttt{cbar} and \texttt{bbar}. For more details on the option of specifying the quark polarisations see Ref. \cite{Kerbizi:2021StringSpinner}.
%The target polarization can be specified by the method \texttt{fhooks.setTargetPol( Vec4 Starget)}, where the spatial part of the four-vector \texttt{Starget} is taken as the target polarization vector in the reference frame chosen for the collision process. Alternatively, the polarization vector of a given quark flavor can be specified using the method \texttt{fhooks.setQuarkPol(int id, Vec4 Squark)}, where \texttt{id} is the \pythia code identifying the quark species and \texttt{Squark} the quark polarization vector in the frame of the collision event (for more details on this option see Ref. \cite{Kerbizi:2021StringSpinner}).

The free parameters of the string+${}^3P_0$ model can be changed in an analogous manner. The complex mass $\mu$ can be specified with the settings
\begin{itemize}\itemsep 0mm
\item \settingval{StringSpinner:re(Mu)}{\textit{reMu}}
\item \settingval{StringSpinner:im(Mu)}{\textit{imMu}}
\end{itemize}
where \textit{reMu} and \textit{imMu} are the values of the real and
imaginary parts of $\mu$ repectively. The parameters which concern
specifically the spin effects for the production and decay of VMs can
be changed by the settings
\begin{itemize}\itemsep 0mm
\item \settingval{StringSpinner:GLGT}{\textit{gLgT}}
\item \settingval{StringSpinner:thetaLT}{\textit{thetaLT}}
\end{itemize}
\textit{gLgT} indicates the value of the coupling constant ratio $|G_{\rm L}/G_{\rm T}|$ which gives the parameter $\fL=|G_{\rm L}/G_{\rm T}|^2/(2+|G_{\rm L}/G_{\rm T}|^2)$. The parameter $\fL$ can be $0\leq\fL\leq 1$. \textit{thetaLT} indicates the value of the parameter $\thetaLT$ which is expected to take values between $-\pi$ and $\pi$. %The complementary methods \texttt{fhooks.getReMu()}, \texttt{fhooks.getImMu()}, \texttt{fhooks.getFL()} and \texttt{fhooks.getThetaLT()} allow instead to access the free parameters. Each method returns a \texttt{double} variable with the value of the corresponding parameter.

As described in detail in Ref. \cite{Kerbizi:2021StringSpinner},
\StringSpinner can produce more general final states allowing the
emission of hadron types that can not be handled by the
string+${}^3P_0$ model, e.g. the baryons and heavier hadronic
states. Their production can be changed using the setting
\settingval{StringSpinner:hadronMode}{\textit{opt}}. Here \textit{opt}
can take the values \texttt{0} or \texttt{-1}. The former value allows
the production of all types of hadrons but the spin effects are
deactivated as soon as the first hadron that can not be handled is
produced. The latter, which is the default setting, rejects all
hadrons that can not be handled in the polarized string fragmentation
(see Sec. \ref{sec:reweight}).

The same options are available also for the treatment of the hadrons that in the string fragmentation process are emitted from the remnant side. This possibility is described in more detail in Ref. \cite{Kerbizi:2021StringSpinner} and the corresponding setting is \settingval{StringSpinner:remnantMode}{\textit{opt}}. The choice \textit{opt=}\texttt{0} activates the splittings from both string ends by propagating the spin effects only from the struck quark whereas the value \texttt{-1}, which is the default setting, rejects the emissions from the remnant side (see Sec. \ref{sec:reweight}).

After the settings of the generator have been specified, the \setting{pythia.init()} command is called followed by the loop over the events. Inside the event loop the same \pythia commands to read off the event record can be used for the analysis of the generated events.

\subsection{Installation and running}\label{sec:execution}
Assuming that \pythia~8.3 has already been installed, the \StringSpinner package can easily be downloaded and installed using the following shell commands
\small
\begin{verbatim}
git clone https://gitlab.com/albikerbizi/stringspinner
cd stringspinner
./configure path/to/pythia/installation/directory
make
\end{verbatim}
\normalsize This produces the file \texttt{Makefile.inc}, which stores
the path to the installation directory of \pythia and is included in
the file \texttt{Makefile}. The latter gathers the necessary commands
to compile the main program.

The main program can then be compiled together with the other files of
the package using the command \texttt{make dis}, which produces the
executable \texttt{dis}, and run with \texttt{./dis}. The cancellation
of the files generated by the compilation procedure can be done using
the \texttt{make clean} command.

%\section{Results from simulations of polarized SIDIS}
\section{Results on the transverse-spin effects in SIDIS}\label{sec:results}

\subsection{Validation and parameter setting}\label{sec:validation}
The new implementation of the spin effects in the \StringSpinner package has been validated by comparing the resulting kinematical distributions of the final state hadrons and the transverse spin asymmetries with those obtained with the standalone implementation of the string+${}^3P_0$ model \cite{Kerbizi:2021M20}. For the comparisons (not shown here) only strings stretched between a transversely polarized $u$ quark and a $(ud)_0$ diquark have been selected, which allows to obtain similar conditions in the two generators. The results obtained with the two generators were compared for different settings of the free parameters and turned out to be consistent.

In this new version of \StringSpinner the setting of the free parameters responsible for the spin effects is chosen as follows. The complex mass $\mu$ is set to $\rm{Re}(\mu)=0.42\,\rm{GeV}/c^2$ and $\rm{Im}(\mu)=0.76\,\rm{GeV}/c^2$, as in Ref. \cite{Kerbizi:2021M20}. The imaginary part of $\mu$ is increased by a factor of two, while keeping the same value of $|\mu|^2$, as compared to the previous version of the package in order to compensate for the dilution of the spin effects arising from the introduction of VMs and their decays in the fragmentation chain. The couplings to the VMs are set to the values $\fL=0.93$ and $\thetaLT=0$, which were shown to give a satisfactory description of the TSA data \cite{Kerbizi:2021M20}. Also the primordial transverse momentum of the quarks in the nucleon is switched off using \settingval{BeamRemnants:primordialKT}{\texttt{off}}. The other parameters are the same as those of \pythia.

In the following subsections we give the main results obtained from simulations of the SIDIS process off transversely polarized protons and compare them with experimental results.

\subsection{Collins asymmetries}
To extract the values of the Collins asymmetries we consider the simulations of the SIDIS process $l p^{\uparrow}\rightarrow l' h X$ off the transversely polarized target proton $p$ at rest. With $l$ and $l'$ we indicate the beam and the scattered leptons, respectively. In the gamma-nucleon reference system (GNS), where the exchanged virtual photon momentum defines the $\zu$ axis and the transverse momentum of the scattered lepton defines the $\xu$ axis, the distribution of the observed hadron $h$ is expected to be \cite{Anselmino:2011ch}
\begin{eqnarray}\label{eq:1h}
    \frac{dN_h}{dx\,dz\,d\pt\,d\phi_C}\propto 1 + \Dnn\,\ST\,\Acoll\,\sin\phiC.
\end{eqnarray}
The distribution is given as a function of the Bjorken variable $x$, the fraction of the exchanged photon energy $z$ carried by $h$, the modulus $\pt$ of the transverse momentum $\ptv$ of $h$, and of the Collins angle $\phiC=\phih+\phiS-\pi$. The angles $\phih$ and $\phiS$ indicate the azimuthal angles of $\ptv$ and of the target transverse polarization $\STv$ in the GNS, respectively. The amplitude of the $\sin\phiC$ modulation is proportional to the product of the depolarization factor $\Dnn=2(1-y)/[1+(1-y)^2]$, $y$ being the fraction of the energy of $l$ carried by the exchanged photon in the GNS, and of the Collins asymmetry $\Acoll(x,z,\pt)$. The possible dependence of the asymmetry on the virtuality $Q^2$ of the exchanged photon is neglected here.

\begin{figure}[tbh]
\centering
\begin{minipage}[b]{0.5\textwidth}
\includegraphics[width=1.0\textwidth]{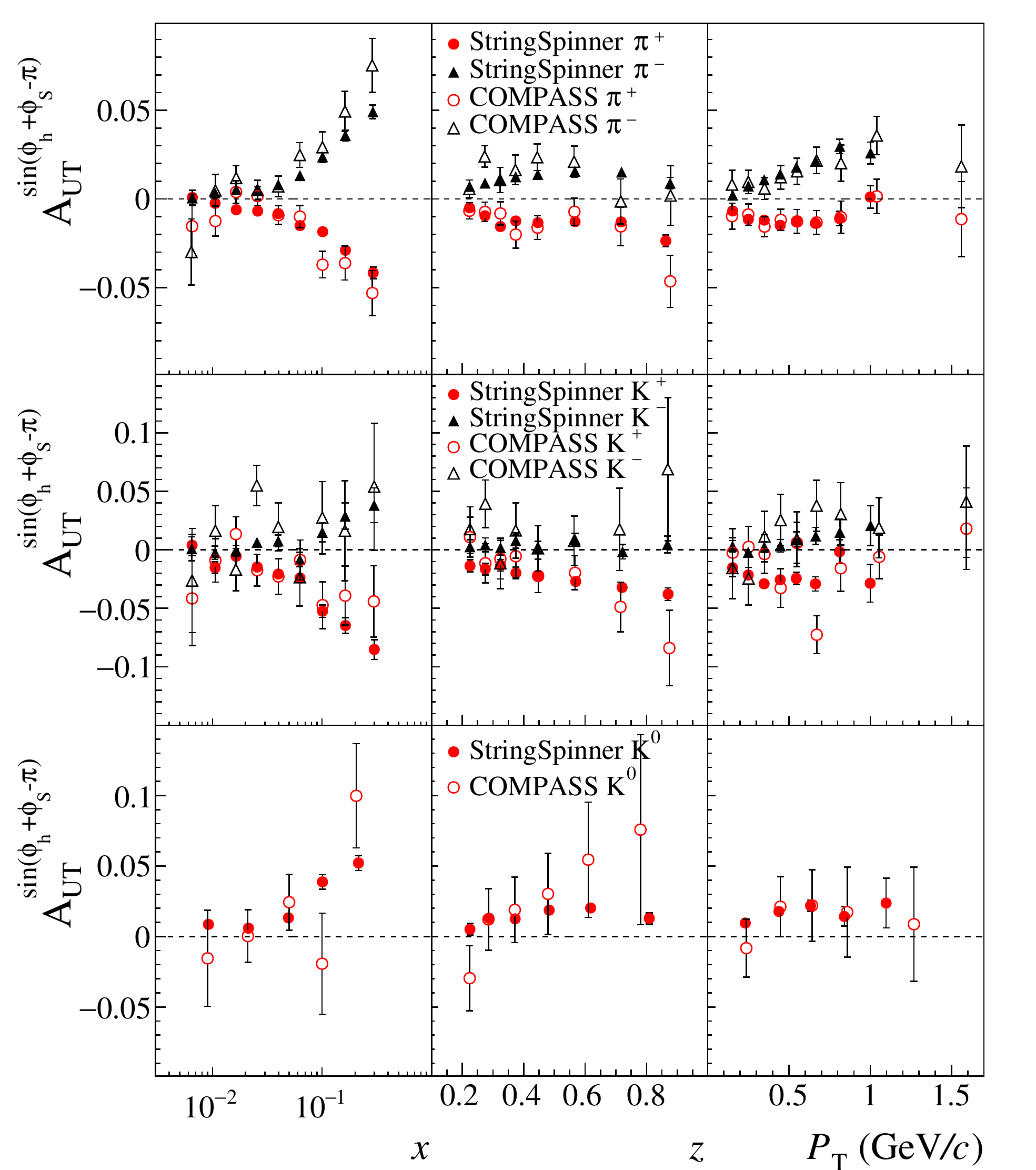}
\end{minipage}
\caption{Comparison between the simulated Collins asymmetries (full points) and the asymmetries measured by COMPASS \cite{COMPASS:2014bze} (empty points) for $\pi^+$ and $\pi^-$ (top row), $K^+$ and $K^-$ (middle row) and $K^0$ (bottom row).}
\label{fig:Collins compass}
\end{figure}

To study the Collins asymmetries in the kinematical configuration of the COMPASS experiment, we carried on simulations of DIS processes with a $160\, \Gevc$ muon beam and applied the same kinematic selections of Ref. \cite{COMPASS:2014bze}. The results are shown as a function of $x$ (left column), $z$ (middle column) and $\pt$ (right column) in Fig. \ref{fig:Collins compass}. The top row gives the asymmetries for $\pi^+$ and $\pi^-$, the middle row those for $K^+$ and $K^-$, and the bottom row that for $K^0$. In each plot the simulation results (full points) are compared with the COMPASS data. As can be seen, \StringSpinner reproduces satisfactorily the trends and the magnitude of the Collins asymmetries for pions. This is also the case for the trend of the $K^+$ asymmetry that, despite the larger uncertainties, show a similarity with the $\pi^+$ asymmetry. In simulations, the $K^+$ asymmetry is somewhat larger as compared to the $\pi^+$ asymmetry owing to the fact that the
kaon sample receives a lower contribution from the decays of VMs. Concerning $K^-$ the measured asymmetry has large uncertainties, which makes the comparison with the trends of the simulation results difficult. The average value of the simulated asymmetry is, however, compatible with that of the data. \StringSpinner reproduces also the sign and the average value of the asymmetry for $K^0$ mesons. An interesting feature of the simulated $K^0$ asymmetry are the large positive values at large $x$, compatible with the data.

\begin{figure}[tbh]
\centering
\begin{minipage}[b]{0.5\textwidth}
\includegraphics[width=1.0\textwidth]{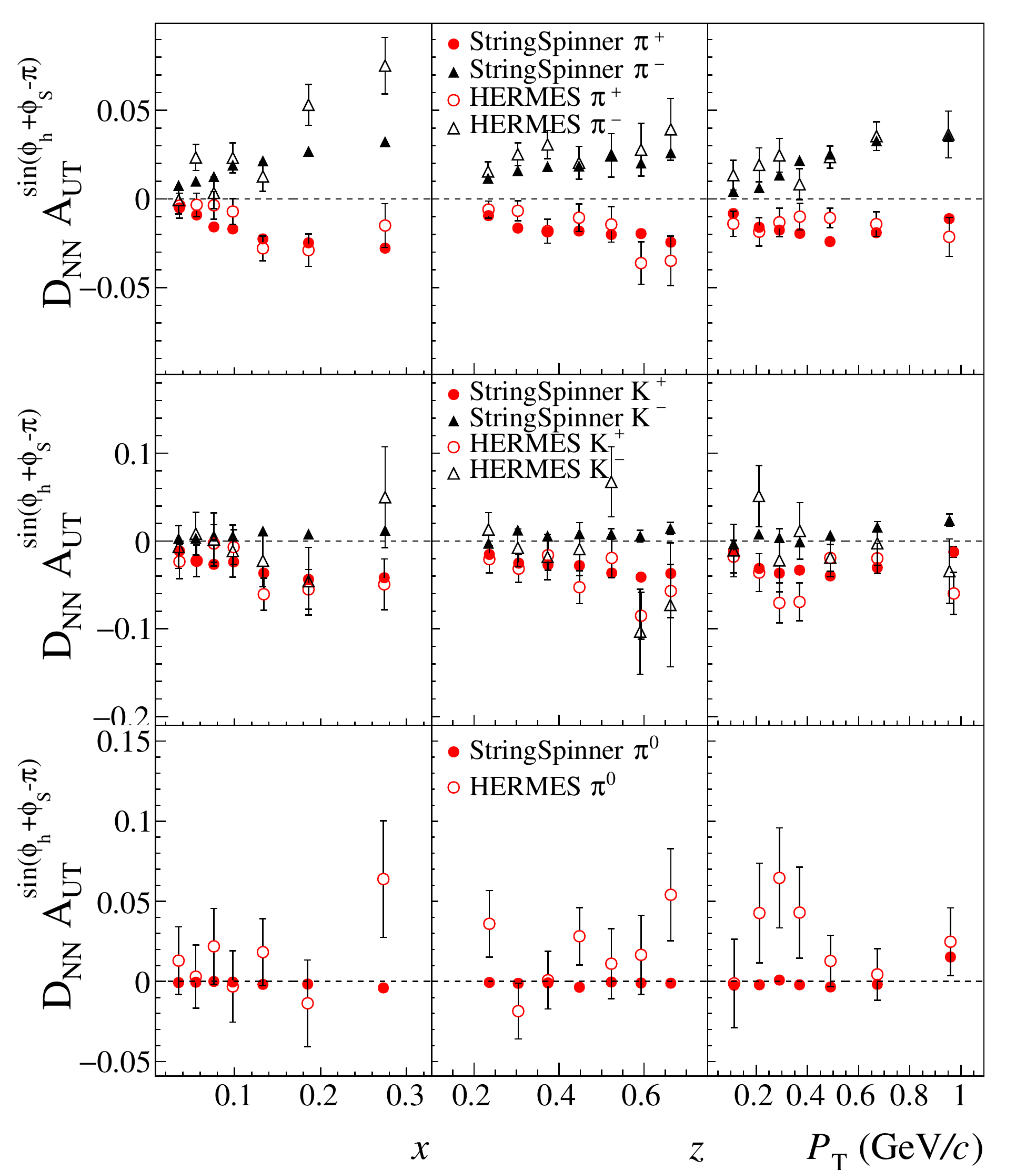}
\end{minipage}
\caption{Comparison between the simulated Collins asymmetries (full points) and the asymmetries measured by HERMES \cite{HERMES:2010mmo} (empty points) for $\pi^+$ and $\pi^-$ (top row), $K^+$ and $K^-$ (middle row) and $\pi^0$ (bottom row).}
\label{fig:Collins Hermes}
\end{figure}

The comparison with the Collins asymmetries from the HERMES experiment \cite{HERMES:2010mmo} is given in Fig. \ref{fig:Collins Hermes}. For this comparison we carried on simulations of DIS processes with a $27.6\,\Gevc$ electron beam and applied the same kinematic selections of Ref. \cite{HERMES:2010mmo}, keeping the same values of the free parameters. The results are shown as a function of $x$ (left column), $z$ (middle column) and $\pt$ (right column). The top row shows the asymmetries for $\pi^+$ and $\pi^-$, the middle row shows the asymmetries for $K^+$ and $K^-$, and the bottom row shows the asymmetry for $\pi^0$ mesons. The simulation results (closed points) are compared to the HERMES data  (open points). Note that the sign of the HERMES asymmetries is reversed in order to use the same convention for the Collins angle as in Eq. (\ref{eq:1h}), and that these asymmetries include the $\Dnn$ factor. \StringSpinner gives an overall satisfactory description of the experimental results, in particular for $\pi^+$, $\pi^-$ and $K^+$. For $x>0.2$ the simulations produce a somewhat lower asymmetry for $\pi^-$ as compared to data but still compatible within the uncertainties. The simulated asymmetry for $K^-$ is small and positive, and compatible with the data within uncertainties. Concerning the asymmetry for $\pi^0$ mesons, the simulations produce a small negative asymmetry as expected by considerations based on isospin symmetry and the fact that the production of $\pi^+$ mesons is favoured in SIDIS off a proton target. The $\pi^0$ asymmetry as measured by HERMES shows instead mostly positive values, although with large uncertainties.

As compared to the Collins asymmetries simulated with the previous version of \StringSpinner, differences can be seen in the trends of the asymmetries as a function of $z$ and $\pt$, as produced by the emission of VMs in the final states and their polarized decays. This has also been studied in detail in the context of the standalone implementation of the string+${}^3P_0$ model in Ref. \cite{Kerbizi:2021M20}. The average values of the Collins asymmetries are instead similar, as expected by the increase of the imaginary part of $\mu$ discussed in Sec. \ref{sec:validation}.

\subsection{The dihadron asymmetry}
%\vspace{-1.5em}
The dihadron asymmetries are obtained from the azimuthal angle distribution of pairs of oppositely charged hadrons $h_1h_2$ produced in the same event. According to the SIDIS cross section for the production of hadron pairs \cite{Bianconi:1999cd}, the distribution of the pair is expected to be
\begin{eqnarray}\label{eq:1hh}
    \frac{dN_{h_1h_2}}{dx\,dz\,d\Mhh\,d\phiRS}\propto 1 + \Dnn\,\ST\,\langle\sin\theta\,\Ahh\rangle\,\sin\phiRS,
\end{eqnarray}
where $z=z_1+z_2$ is the fractional energy of the pair and $\Mhh$ is its invariant mass. The azimuthal angle $\phiRS$ is measured in the GNS and it is defined to be $\phiRS=\phiR+\phiS-\pi$. Here $\phiR$ is the azimuthal angle of the relative transverse momentum of the pair $\RT=z_2\,\ptOne/z-z_1\ptTwo/z$, where $\ptOne$ and $\ptTwo$ are the transverse momenta of the hadron $h_1$ (the positive one) and $h_2$ in the GNS, respectively. The dihadron asymmetry $\langle \sin\theta\,\Ahh\rangle$ appears in the amplitude of the $\sin\phiRS$ modulation and it can depend on the variables $x$, $z$, and $\Mhh$. The angle $\theta$ is the polar angle of one of the two hadrons measured in the rest frame of the pair with respect to the direction of the total momentum of the pair. The angle brackets indicate the integration w.r.t.\ this angle.

\begin{figure}[tbh]
\centering
\begin{minipage}[b]{0.5\textwidth}
\includegraphics[width=1.0\textwidth]{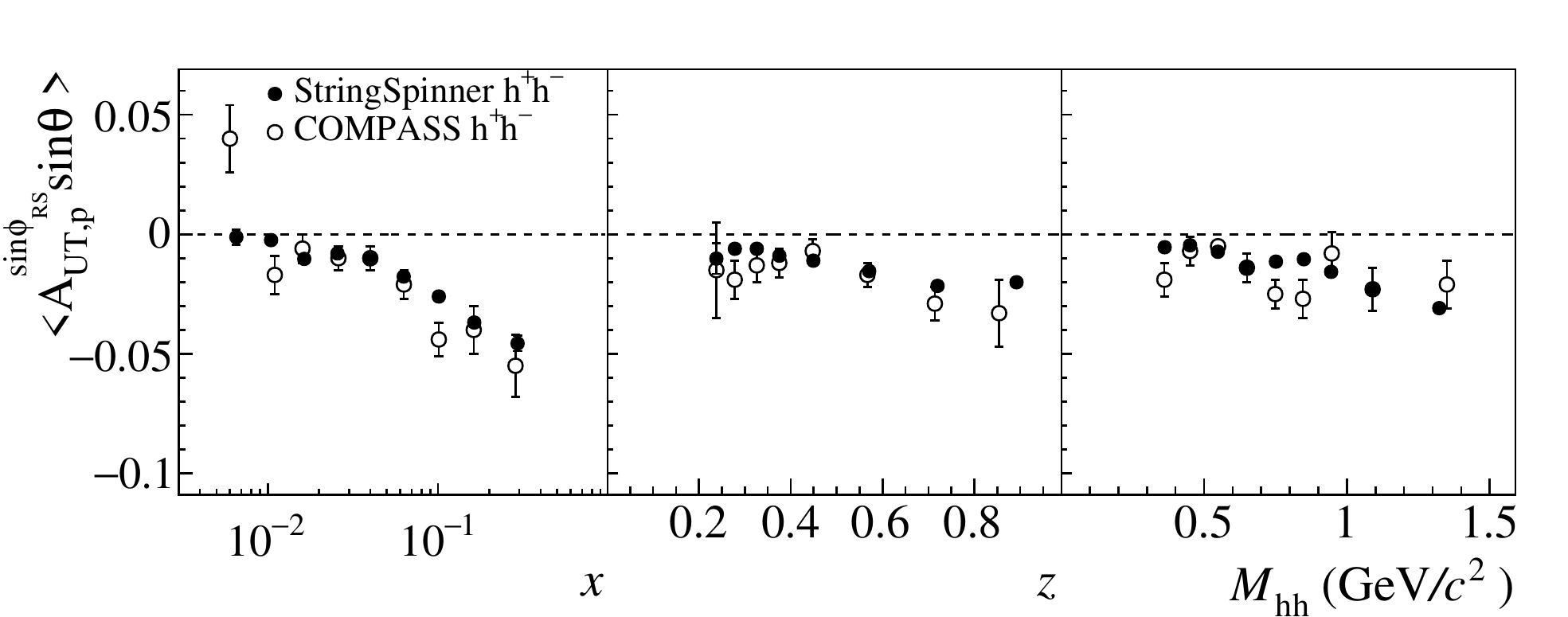}
\end{minipage}
\caption{Comparison between the simulated dihadron asymmetries (full points) and the asymmetries measured by COMPASS \cite{COMPASS:2014ysd} (empty points) for $h^+h^-$ pairs.}
\label{fig:Dihadron compass}
\end{figure}

The simulated dihadron asymmetry for charged hadron pairs, namely $h_1=h^+$ and $h_2=h^-$, obtained from simulations of the SIDIS process off a transversely polarized proton at rest in the COMPASS kinematics, is shown in Fig. \ref{fig:Dihadron compass} (open points) as a function of $x$ (left plot), $z$ (middle plot) and $\Mhh$ (right plot). It is compared with the dihadron asymmetry as measured by COMPASS \cite{COMPASS:2014ysd} (open points). In the simulations the same selection cuts as in the COMPASS analysis are applied. The description of the dihadron asymmetry by \StringSpinner is satisfactory for all kinematic variables. Looking at the asymmetry as a function of $\Mhh$, a somewhat different trend can be seen in the simulations in the invariant mass region of the $\rho^0$ meson. While the simulated asymmetry is diluted by the $\rho^0$ decays, the data seem to indicate a possible enhancement. This could be due to possible interference effects among the pion pairs produced in the $\rho^0$-decay and those produced directly from string fragmentation, which are not considered in the simulations.

Compared to the previous version of \StringSpinner, the trend of the dihadron asymmetries simulated with the new version as a function of invariant mass is somewhat different. It is due to the fact that the decays of VMs contribute only to the dilution of the dihadron asymmetry in Eq. (\ref{eq:1hh}) \cite{Kerbizi:2021M20}, which results in structures in the invariant mass regions of the VMs.

\subsection{Collins asymmetries for vector mesons}
The Collins asymmetry for VMs is an important observable to access information on the spin-dependence of the hadronization process, as discussed in detail in Ref. \cite{Kerbizi:2021M20}. The Collins asymmetries for VMs can be extracted from the simulated data using Eq. (\ref{eq:1h}) and selecting the meson among the produced hadrons in the event. From the experimental point of view, however, the measurement of the Collins asymmetries for VMs is challenging. The main reason is the large combinatorial background from non-resonant meson pairs that must be subtracted when constructing the candidate VMs. The data on the Collins asymmetries for VMs is presently poor, the only existing measurement being the Collins asymmetry for $\rho^0$ mesons measured in SIDIS off protons by COMPASS \cite{COMPASS:2022jth}.  %In fact, according to simulations carried with \pythia, vector mesons are mainly produced directly from string fragmentation and the contribution from the decay of heavier resonances is expected to be small. This means that vector mesons carry in principle a clean information on the production mechanisms at play in string fragmentation. Also, according to the string+${}^3P_0$ model, transverse-spin effects for the production of VMs are expected to be sensitive to their polarization states as shown in Ref. \cite{Kerbizi:2021M20} for the case of Collins asymmetries.

\begin{figure}[tbh]
\centering
\begin{minipage}[b]{0.5\textwidth}
\includegraphics[width=1.0\textwidth]{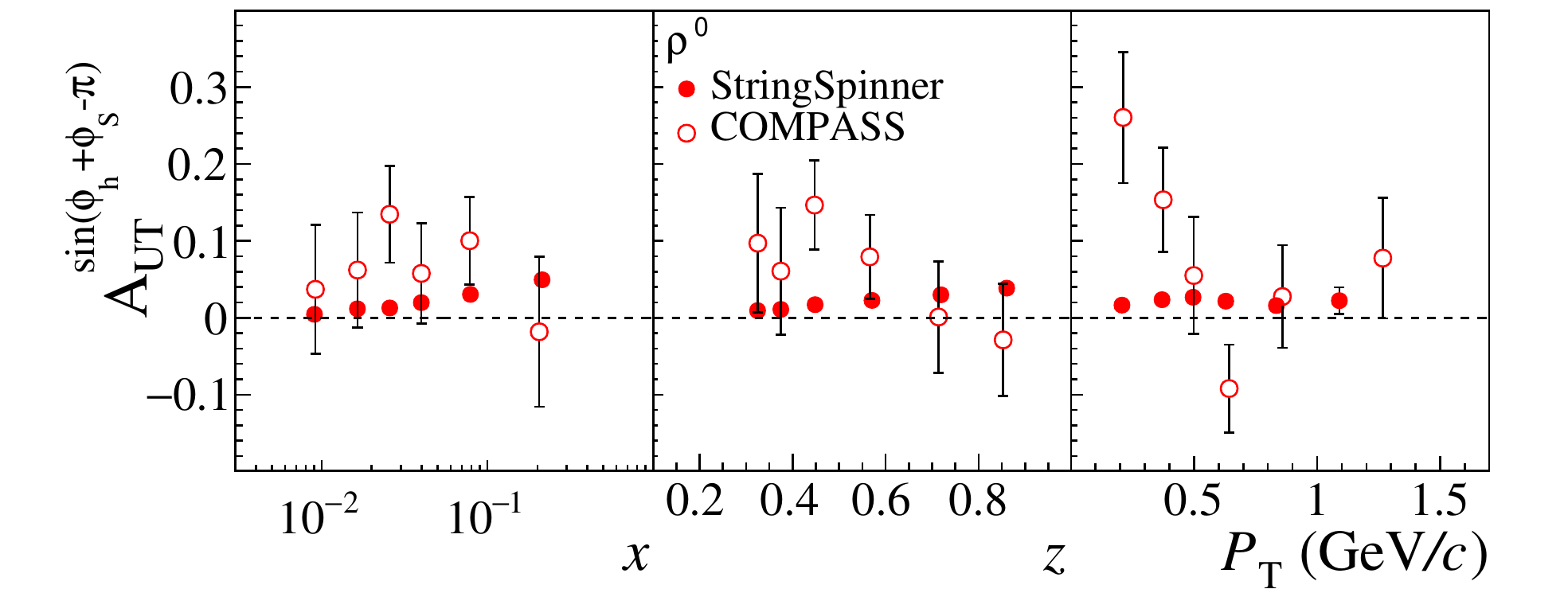}
\end{minipage}
\caption{Comparison between the simulated Collins asymmetry for $\rho^0$ mesons (full points) and the corresponding asymmetry measured by COMPASS \cite{COMPASS:2022jth} (empty points).}
\label{fig:rho0 compass}
\end{figure}

The comparison between the simulated Collins asymmetry for $\rho^0$ mesons with the COMPASS data is shown in Fig. \ref{fig:rho0 compass}. We stress that this is the first time that the simulation of the full Collins asymmetry for $\rho^0$ mesons (and in general for that of VMs) produced in SIDIS is made possible. The asymmetry is shown as a function of $x$ (left plot), $z$ (middle plot) and $\pt$ (right plot). The same selection cuts as in the COMPASS analysis are applied also in the simulations. The simulated asymmetry (full points) shows positive values as in the data (open points). As expected from the string+${}^3P_0$ model \cite{Czyzewski:1996ih,Kerbizi:2021M20}, the Collins asymmetry for $\rho^0$ mesons has the opposite sign as compared to the asymmetry for $\pi^+$ mesons (see Fig. \ref{fig:Collins compass}). %The large statistical uncertainties of the data do not allow a comparison with the kinematic dependencies of the asymmetry.
The simulation results exhibit an increasing trend as a function of $x$ and $z$, and a decreasing trend for $\pt<0.5\,\rm{GeV/c}$. The data give indications for positive asymmetries as well, but the large statistical uncertainties make a comparison of the dependence on the kinematic variables difficult.
%Despite the large uncertainties in the data, this is the only existing measurement indicating a positive sign of the Collins asymmetry for $\rho^0$ mesons.

\section{Conclusions}\label{sec:conclusions}
We presented the new version of the \StringSpinner package that includes the production of vector mesons and their polarized decays in the string fragmentation process for DIS events. For this purpose we used the string+${}^3P_0$ model of polarized hadronization, whith pseudoscalar and vector meson production. The main modifications to the previous version of the package concern the production of vector mesons in the final state, the simulation of the external decay processes of such mesons and the interface to the most recent \pythia~8.3 event generator. The developments presented here provide in addition the framework for possible future extensions of the package such as the production of other hadron species in the polarized quark fragmentation and for its use in simulating other collision processes.

The new package was used to simulate transverse spin asymmetries in SIDIS, which were compared with the experimental data showing a satisfactory agreement. In particular, the results for the Collins asymmetries agree with the measured ones for different species of hadrons, i.e. charged and neutral pions and kaons, and $\rho^0$ mesons.

The encouraging results obtained in this work confirm the soundness of the string+${}^3P_0$ model in the description of the (transverse) spin effects in hadronization, and the framework presented here is a promising candidate for the systematic introduction of spin effects in the \pythia event generator. It would be a valuable tool
to be used in the preparation of future experiments like the EIC, and deserves further development.



\section*{Declaration of competing interest}
The authors declare that they have no known competing financial interests or personal relationships that could have appeared to
influence the work reported in this paper.

\section*{Acknowledgement}
AK is grateful to Prof. Anna Martin and Prof. Xavier Artru for the many discussions and the valuable comments on the topic of this work. He also acknowledges partial support by NextGenerationEU within the Piano Nazionale di Ripresa e Resilienza (PNRR) - CUP n. J97G22000510001. Part of the work of AK was done in the context of the POLFRAG project.

%% References


\bibliographystyle{elsarticle-num}
%

%\bibliography{bibliography.bib}






% \end{thebibliography}

\end{document}